\documentstyle[prl,aps,epsfig,floats]{revtex}

\def\qd{{\bf Q}_{\delta}}
\def\qaf{{\bf Q}_{\rm AF}}
\def\lsno{La$_{2-x}$Sr$_{x}$NiO$_{4}$}
\def\lno{La$_{2}$NiO$_{4}$}

\begin{document}
\title{High-Energy Spin Dynamics in La$_{1.69}$Sr$_{0.31}$NiO$_4$}
\author{ P. ~Bourges$^1$,  Y. ~Sidis$^1$, M. Braden$^{1}$, K. Nakajima$^2$,
and J.M. Tranquada$^3$}
\address{
$^1$ Laboratoire L\'eon Brillouin, CEA-CNRS, CE-Saclay, 91191 Gif sur
Yvette, France\\
$^2$ Neutron Scattering laboratory, ISSP, University of Tokyo, Tokai, Ibaraki,
Japan\\
$^3$ Physics Department, Brookhaven National Laboratory, Upton, NY 11973, USA\\
}
\date{\today}

\twocolumn[\hsize\textwidth\columnwidth\hsize\csname@twocolumnfalse\endcsname

\maketitle

\begin{abstract}

We have mapped out the spin dynamics in a stripe-ordered nickelate,
\lsno\ with $x \simeq 0.31$, using inelastic neutron
scattering.  We observe spin-wave excitations up to 80 meV emerging from the
incommensurate  magnetic peaks with an almost isotropic spin-velocity:
$\hbar c_s\sim 0.32$ eV \AA, very similar to the velocity in the undoped,
insulating parent compound, \lno.  We also discuss the
similarities and differences of the inferred spin-excitation spectrum with
those reported in superconducting high-$T_c$ cuprates.

\end{abstract}

\pacs{PACS numbers: }
]

\narrowtext

Magnetism plays an important role in several theories of the high-temperature
superconductivity found in layered cuprates; hence, experimental
characterizations of magnetic excitations in these materials have been of
considerable interest.  Much attention has been focussed on the ``resonance''
peak observed by inelastic neutron scattering in a number of different
cuprates \cite{he02}.  The resonance peak is centered commensurately on
the antiferromagnetic wave vector.  Studies of YBa$_2$Cu$_3$O$_{6+x}$, in
which the resonance peak was first observed \cite{bour98}, have also found
incommensurate excitations at somewhat lower energies \cite{mook98,arai99}.  In
more recent work \cite{bour00}, it was found that the incommensurate
scattering actually disperses downward continuously from the commensurate
resonance peak, defining a single dispersive excitation.

There are several theoretical perspectives on the resonance peak and the
dispersive excitations.  In one popular approach, the magnetic resonance is a
particle-hole bound state below the two-particle continuum associated with the
$d$-wave superconducting gap \cite{onuf02,norm01,chub01,brin01};
such calculations, based on a homogeneous, renormalized Fermi-liquid model,
yield qualitative agreement with experiment.  One alternative is based on the
``stripe'' scenario \cite{emer99,zaan99}, in which magnetic excitations are
dominantly attributed to spatially segregated domains of antiferromagnetically
correlated copper spins \cite{zaan96a,hass99,sach00,kane01,varl02}.  In
particular, Batista, Ortiz, and Balatsky \cite{bati01} have explicitly
proposed that the dispersive resonance represents the magnon-like
excitations emanating from incommensurate wave vectors associated with a
stripe-correlated spin system.  Their prediction that a similar resonance-like
excitation should be observable in a stripe-ordered compound such as \lsno\
motivated the present investigation.

Regardless of whether \lsno\ is an ideal model for the cuprates, studies of
the full spin dynamics of the incommensurate spin state are of interest, as only
low-energy characterizations have been reported previously
\cite{tran97c,lee02}. Here we report measurements of the high-energy spin
excitations in a crystal with $x\simeq 0.31$, close to the $1/3$ composition.
Within the two-dimensional reciprocal space corresponding to a doped NiO$_2$
plane, the diagonal stripe order yields two pairs of magnetic ordering wave
vectors, $\qd=(\frac12,\frac12)\pm(\delta,\delta)$ and
$(\frac12,\frac12)\pm(\delta,-\delta)$, due to twinning of the stripe
domains.  [We express wave vectors in units of the reciprocal lattice,
$(2\pi/a,2\pi/a)$, with $a=3.82$~\AA.] These incommensurate points
are displaced about the antiferromagnetic propagation wave vector,
$\qaf=(\frac12,\frac12)$, of the undoped parent compound.  The
incommensurability $\delta$ varies with the hole concentration as
$\delta\approx x/2$ \cite{yosh00}, with $\delta=0.158$ for our sample at
low temperature.  We observed spin-wave excitations dispersing from each
$\qd$ peak up to a maximum of 80 meV at $\qaf$.  Surprisingly, the
effective spin-wave velocity, $\hbar c_s\sim0.32$~eV~\AA, is almost as large as
that of undoped insulating \lno, where $\hbar c_0=0.34$ eV~\AA\cite{naka93}.
Furthermore, the anisotropy in the spin-wave velocity between the directions
parallel and perpendicular to the stripes is less than 15\%.  On warming
to above the charge-ordering temperature, the excitation at $\qaf$
softens somewhat, but remains well-defined.

Our single-crystal sample,
grown by the floating-zone method, was the subject of a previous neutron
scattering experiment \cite{tran02}.  The present
inelastic-neutron-scattering measurements were performed on the 1T and 2T
triple-axis spectrometers at the Orph\'ee reactor of the Laboratoire
L\'eon Brillouin in Saclay, France.  Each spectrometer is equipped with
Cu (111) and pyrolytic graphite (PG) monochromators and a  PG (002)
analyzer.  Different final neutron energies of the analyzer, $E_f=14.7$,
30.5 and 41 meV, were used in order to cover the full energy range of the
magnetic spectrum.   A PG filter was placed after the sample to minimize
neutrons at higher-harmonic wavelengths.  Most of the measurements have
been performed within the $(HK0)$ scattering plane, with some data
collected in the $(HHL)$ zone.

\begin{figure}[t]
\epsfxsize=8 cm
$$
\epsfbox{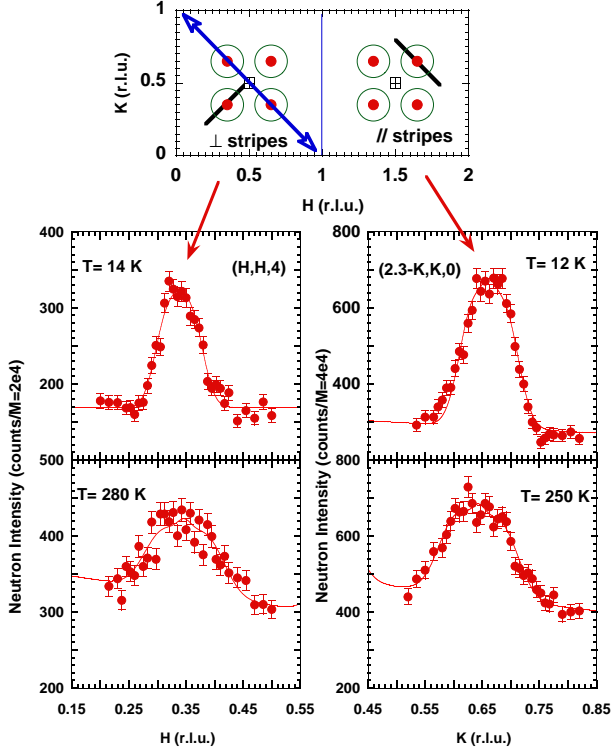}
$$
\caption{ 
upper panel) Sketch of the Brillouin zone in the reciprocal $(HK0)$-plane.
Full red circles indicate the location of the incommensurate magnetic
peaks, $\qd$. The green circles show the cut of the spin-waves cones at a
constant energy, defining $Q_\omega$ ($\equiv \omega/c_s$ in the
low-energy regime). Black bars represent the directions of the scans
shown in the lower panel, the blue arrow shows the direction of the scans
shown in Fig. \ref{qscan_disp}. lower panels) Constant energy scans at
$E=28.1$~meV (measured with $E_f= 14.7$ meV) at two temperatures along
the two directions sketched in the upper panel, probing fluctuations
along wave vectors perpendicular (left) and parallel (right) to the
stripes.  Lines represent best fits of the extended Gaussian model
described in the text. }
\label{28meV}
\end{figure}

\begin{figure}[t]
\vspace*{-0.5cm}
\epsfxsize=6cm
$$
\epsfbox{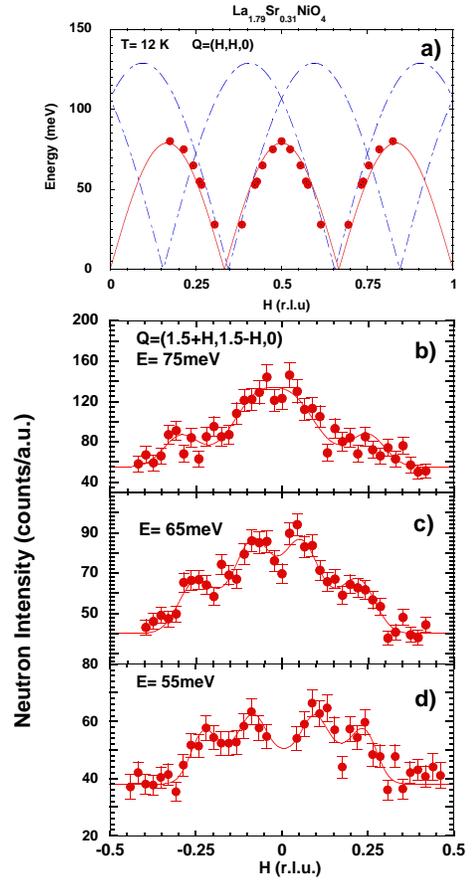}
$$
\caption{ 
Upper panel) Dispersion relation of the spin excitations in
La$_{1.69}$Sr$_{0.31}$NiO$_4$.  The full red line is a fit by a simple
$|\sin(3\pi H)|$ function. The two blue dashed lines correspond to the
spin-wave dispersion relation in undoped \lno, but shifted to the
incommensurate wave vectors. Lower panels) Constant energy scans at b) $E=75$
meV, c) $E=65$ meV and d) $E=55$ meV along $(H,-H,0)$ (i.e., perpendicular to
the stripes) around $Q=(1.5,1.5,0)$. (b) was measured with $E_f=41$ meV (as
in Fig.~\ref{energy}), while (c) and (d) were measured with $E_f=30.5$ meV.
Full lines represent best fits of the extended Gaussian model described in the
text.}
\label{qscan_disp}
\end{figure}

In scanning across the $\qd$ peak positions at constant energy
transfer, as indicated in the upper panel of Fig. \ref{28meV}, we expect, at
low temperatures (below the spin-ordering temperature of 160~K), to pass
through two spin-wave branches, corresponding to counter-propagating
excitations. The lower panels show scans at an energy transfer of 28.1~meV
along orthogonal directions with respect to the modulation wave vector.
Although the two branches are not resolved, the $Q$-width in each case is
significantly broader than resolution, with little dependence on scan
direction.  At higher energies [Fig.~\ref{qscan_disp}(b)-(d)], we clearly
resolved spin waves emerging from the two $\qd$ points
$(1.5-\delta,1.5+\delta)$ and $(1.5+\delta,1.5-\delta)$.  At $E=75$~meV,
the excitations from $+\delta$ and $-\delta$ begin to merge at $\qaf$,
forming a single broad peak, with additional weight
coming from excitations associated with the orthogonal stripe domains (see
Fig.~\ref{28meV}).

\begin{figure}[t]
\vspace*{0cm}
\epsfxsize=6cm
$$
\epsfbox{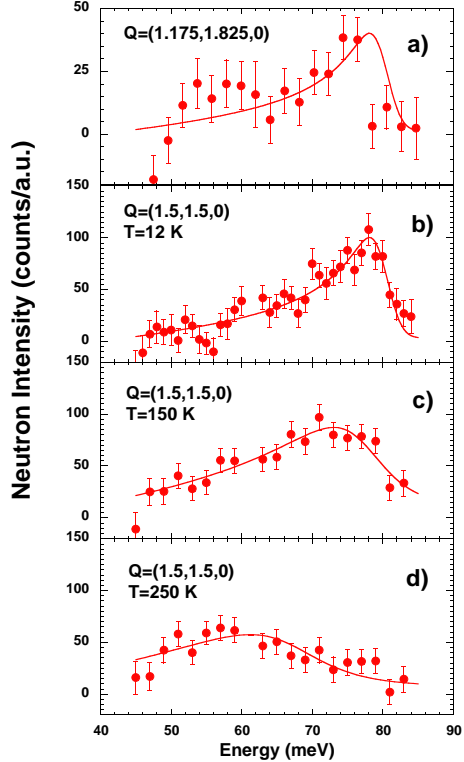}
$$
\caption{ 
Energy scans at (a) ${\bf Q}=(1.175,1.825,0)$, (b)-(d)
${\bf Q}=(1.5,1.5,0)\equiv\qaf$, after subtraction of background measured
either at ${\bf Q}=(1,2,0)$ or by rocking the sample by $\pm15^\circ$.
The data have been obtained with $E_f=41$ meV.
The full lines represent best fits of a damped
harmonic oscillator function, with a dispersion relation as shown in
Fig.~\ref{qscan_disp}(a), convolved with the  spectrometer resolution. The line
in (a) is the same as in (b) except for a scale factor.}
\label{energy}
\end{figure}

The excitation at $\qaf$ is better characterized by scanning the energy
transfer at fixed momentum transfer.  Figure~\ref{energy}(b)-(d) shows energy
scans at $\qaf$ for 3 different temperatures.  The maximum of the spin
excitation spectrum is $\omega_0 = 80 \pm 0.8$  meV
at low temperature [Fig.~\ref{energy}(b)]. 

In the studies of low-energy excitations in stripe-correlated
nickelates \cite{tran97c,lee02}, it was found that the observed
$Q$-widths of the excitation peaks are broader than the energy
resolution.  For Sr-doped samples, this is due, at least in part, to the
finite spin-spin correlation length in the ordered state.  Fluctuations
of the charge stripes might also play a role.  To extract the frequency
dispersion from the constant-energy scans of Figs.~\ref{28meV} and
\ref{qscan_disp}, we have assumed a scattering function of the form
\begin{equation}
    S({\bf Q},\omega) = A f^2(Q) \sum_{domains}  \sum_\delta
    e^{-[({\bf Q}-\qd)^2-Q_\omega^2]/2\Delta^2},
   \label{eq_exp}
\end{equation}
where $A$ is a scale factor, one sum is over the two incommensurate wave
vectors,  the other one is  over the twined stripe domains, and
the magnetic form factor $f(Q)$ is assumed to vary insignificantly
across the range of a given scan.  To fit the data, the model $S({\bf
Q},\omega)$ was convolved with the spectrometer resolution function, and
the parameters $Q_\omega$ and $\Delta$ were varied to minimize $\chi^2$.
The model provides reasonable fits to the data; note that the enhanced
intensity near $\qaf$ in Fig.~\ref{qscan_disp}(b) and (c) comes from the
superposition of contributions dispersing from the four surrounding $\qd$
points (see the upper panel of Fig.~\ref{28meV}).   The results obtained
for $Q_\omega$ as a function of energy transfer, $\hbar\omega$, are
plotted in Fig.~\ref{qscan_disp}(a). The momentum width, $\Delta$, when
converted to half-width-at-half-maximum, is 0.05~\AA$^{-1}$.

\begin{figure}[t]
\epsfxsize=8cm
$$
\epsfbox{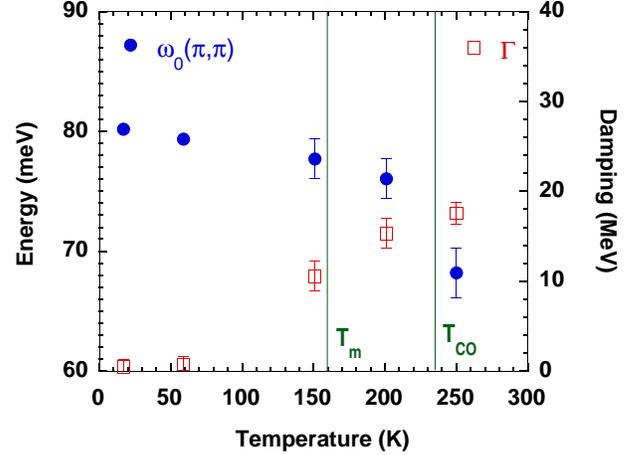}
$$
\caption[xxx]{ 
Temperature dependence of various fitting parameters:  Spin-wave energy maximum
(left scale) and damping energy at $\qaf$ (right scale).  Vertical lines
indicate the magnetic and charge-order transitions.}
\label{temperature}
\end{figure}

To fit the constant-{\bf Q} scans of Fig.~\ref{energy}, it is standard to
use an alternative parametrization:
\begin{equation}
    S({\bf Q},\omega) = A' f^2(Q)
    {\omega\Gamma [1+n(\omega,T)] \over[\omega^2-\omega^2({\bf Q})]^2 +
\omega^2\Gamma^2},
\end{equation}
where $\omega({\bf Q}) = \omega_0-\alpha({\bf Q}-\qaf)^2$ ($\alpha= 170$
meV \AA$^2$), and $n(\omega,T)$ is the
Bose temperature factor.  The temperature-dependence of the 
parameters $\omega_0$
and $\Gamma$ are plotted in Fig.~\ref{temperature}, while the
low-temperature result  for $\omega_0$ is plotted in
Fig.~\ref{qscan_disp}(a).

The spin-wave dispersion determined by the quantitative analysis can be
described fairly well by a very simple expression:
$\omega({\bf Q})=\omega_0|\sin(3\pi H)|$ along the $[HH0]$ direction,
indicated by the full line in Fig.~\ref{qscan_disp}(a).  For comparison,
the spin excitation spectrum measured in undoped  \lno\ \cite{naka93},
shifted from
$\qaf$ to $\qd$ is indicated by dot-dashed lines. It matches the low-energy
behavior surprisingly well, but clearly deviates at high energy.  The $\qaf$
crossing of the shifted curves is at 106 meV, while the measured $\omega_0$ is
renormalized down to 80 meV.  Besides the maximum at $H=\frac12$, the model
dispersion curve also has maxima at $H=\pm\frac16$.  The energy
scan in Fig. \ref{energy}(a) shows that the observed maximum energy at
$H\approx\frac16$ is essentially the same as that at $\qaf$, and that the
structure factors are similar as well. Looking at Fig.~\ref{qscan_disp}(a), one might 
expect to find an additional optical spin-wave branch at higher energies ($\alt$ 125 meV); 
searches at energies of up to 100 meV did not yield any positive evidence 
for such a branch.

It is interesting to compare the doping dependence of the maximum
spin-excitation frequency, $\omega_0$, with that of the ``2-magnon frequency'',
$\nu_{\rm 2mag}$, determined by Raman scattering \cite{blum98,yama98,suga98}.
We find that the ratio
   $\omega_0(x=0.31)/\omega_0(x=0)=80~{\rm meV}/124~{\rm meV}=0.65$
is very similar to
$\nu_{\rm 2mag}(x=0.33)/\nu_{\rm 2mag}(x=0)= 1110~{\rm 
cm}^{-1}/1640~{\rm cm}^{-1}
    =0.68$.
In the Raman studies of the $x=0.33$ phase \cite{blum98,yama98}, a second,
lower-energy peak at 720~cm$^{-1}$ ($\equiv$ 90 meV) was also 
attributed to 2-magnon scattering;
however, we do not observe any features in the single-magnon dispersion that
would correlate with a second 2-magnon peak.  Alternatively, a Raman study of
oxygen-doped \lno\ \cite{suga98} suggests that the 720~cm$^{-1}$ feature might
be associated with a phonon mode, observed even in undoped \lno, that should
be Raman inactive.

To evaluate the anisotropy in the dispersion, we return to the 28-meV data of
Fig.~\ref{28meV}.  By fitting the resolution-convolved Eq.~(\ref{eq_exp}) to
the low temperature data, we can estimate the spin wave velocity,
$c_s=\omega/Q_\omega$, along directions parallel and
perpendicular to the stripes.  Consistent with the figure, we find little
anisotropy, with $\hbar c_{s\|}=300\pm20$ meV \AA\ and
$\hbar c_{s\bot}=350\pm20$ meV \AA. Both of these values are close to the value
of $\hbar c_0=340$ meV \AA\ obtained in pure \lno \cite{naka93}.
The lack of significant anisotropy is rather counter-intuitive.
The observation of
well-defined excitations  following a simple dispersion curve such as shown in
Fig.~\ref{qscan_disp}(a) is rather remarkable relative to the large
hole concentration. One can suggest that charge order
coherently affects the spin dynamics for this hole filling near
${1\over3}$ due to 
pinning effects.

As shown by Figs.~\ref{28meV} and \ref{energy},
a striking temperature  dependence is observed at energies much larger than the
temperature. With increasing temperature, the magnetic mode at $\qaf$
softens and broadens upon heating, starting near the magnetic ordering
temperature, $T_m$ (Fig.~\ref{temperature}).  The frequency reduction and
damping are of comparable magnitude.  It is of particular interest that the
mode remains underdamped above the charge-ordering transition.  This
observation is consistent with the stripe-liquid phase proposed to describe
the low-energy spin dynamics\cite{lee02}.

Returning to our original motivation \cite{bati01}, the spin excitation
spectrum in our stripe-ordered nickelate does have a straightforward
similarity to that measured in the superconducting state of cuprates
\cite{bour00}. In both cases, a downward dispersion is observed below a
maximum frequency at the AF wave vector.  Of course, there are also clear
differences: in particular, the spin excitation spectrum of the
nickelate is not limited to only a small portion of reciprocal space
around $\qaf$. We observe symmetric dispersions and structure factors
about the incommensurate  wavevectors. These results compare somewhat
better with the excitations observed in La$_{2-x}$Sr$_x$CuO$_4$,  where
low energy incommensurate peaks appear to merge into a broad commensurate
response around $\sim 25$ meV \cite{peti97,mats94}; however, in the
latter case the merging occurs at a much lower energy than the value
one would deduce from the cuprate spin-wave velocity and the
incommensurability. In any case, the canonical nature of the ordered
stripes in the nickelate limits the degree to which a direct comparison
with the cuprates is practical. A more relevant point for comparison may
be the high-temperature phase of the nickelate where, in the absence of
static magnetic  or charge order, the high-energy spin excitations remain
underdamped.

In conclusion, we have measured the complete spin excitation spectrum in
the stripe-ordered nickelate  La$_{1.69}$Sr$_{0.31}$NiO$_4$.  The deduced
spin-wave velocity is surprisingly large, and there is remarkably little
anisotropy between directions parallel and perpendicular to the stripes.
This spin dynamics study, in comparison with those reported for
superconducting cuprates, should help to clarify the role and relevance
of stripes in oxides.

We wish to thank A.V. Balatsky, C. Morais-Smith, and J. Zaanen for
fruitful  discussions.  Work at Brookhaven is supported by U.S.
Department of Energy Contract No.\ DE-AC02-98CH10886. Work at ISSP is supported 
by a Grant-In Aid for Scientific Research from the Ministry of Education, Culture, 
Sports, Science, and Tchnology, Japan


\end{document}